# A risk-based approach to assessing liability risk for AI-driven harms considering EU liability directive

Self-assessment of liability risk for AI-driven harms

Sundaraparipurnan N,

University of Bordeaux, sundar.narayanan@aitechethics.com

Mark Potkewitz, BROOKLYN LAW SCHOOL, mark.potkewitz@brooklaw.edu

Abstract

Artificial intelligence (AI) can cause inconvenience, harm, or other unintended consequences in various ways, including those that arise from defects or malfunctions in the AI system itself or those caused by its use or misuse. Responsibility for AI harms or unintended consequences must be addressed to hold accountable the people who caused such harms and ensure that victims are made whole concerning any damages or losses they may have sustained. Historical instances of harm caused by AI have led to the EU establishing an AI liability directive. This directive aims to lay down a uniform set of rules for access to information, delineate the duty and level of care required for AI development and use, and clarify the burden of proof for damages or harms caused by AI systems, establishing broader protection for victims. The future ability of the provider to contest a product liability claim will rely on good practices adopted in designing, developing, deploying, and maintaining AI systems in the market. This paper provides a risk-based approach to examining liability for AI-driven injuries. It also provides an overview of existing liability approaches, insights into limitations and complexities in these approaches, and a detailed self-assessment questionnaire to assess the risk associated with liability for a specific AI system from a provider's perspective.

## 1 INTRODUCTION

Using AI alongside other technologies, we could enable machines to make decisions faster than humans and perform actions more quickly. AI-based systems can also make more objective, consistent, and trustworthy decisions than humans.

Artificial intelligence (AI), which has the potential to generate value and create risks, has seen its potential exponentially increase in recent years. AI systems' characteristics include complexity, transparency, and autonomy during operation. Currently, AI systems operate autonomously and will continue to do so. This means that it is essential to provide incentives for human participants in AI systems so they can reduce AI-related harm [1].

Artificial intelligence can cause harm in various ways, including those arising out of defects or malfunctions in the AI system itself or the unintended consequences of its use. Without the appropriate safeguards, AI can exercise or exacerbate racial or gender-and-sex-based biases, social discrimination, human job losses, and, in extreme cases, physical harm. Responsibility for AI injuries and harms is essential to ensure that the people who caused them are held accountable and that victims receive compensation for any damages or losses they may have sustained. Historical instances of harm caused by AI have led to the European Union (EU) establishing an AI liability directive. Manufacturers and sellers may not be motivated to ensure that AI products are reliable and safe for consumers or end-users. Victims may only have limited recourse to seek compensation.

The EU Directive aims to lay down uniform rules for access to information and alleviate the burden of proof for damage or harms caused by AI systems, establishing broader protection for victims.

This article attempts to create a self-assessment checklist to assess liability risk for AI products. The guidelines cover liability applicability, proof burden, and liability exemptions. With more and more questions arising, legal and regulatory frameworks will evolve and adapt to the changing need, including having more specific mechanisms for liability determination than what exists today [2]. This article attempts to bring a risk-based approach to managing complexity relating to liability and provides a detailed self-assessment questionnaire to determine the liability risk of the AI system in question from a provider's perspective. This paper establishes the broader context of harms with specific examples in Section 2, uniqueness regarding liability for AI harms and the complexities associated with liability assessments in Section 3, critical insights from the European Union (EU) AI liability directive in Section 4, complexities related to liability assessment in Section 5, a risk-based approach to liability, and an illustrative questionnaire in Section 6.

## 2 AI-DRIVEN HARMS IN THE REAL WORLD

AI-driven harms are any harms or losses caused by AI, its use, or the system itself. This harm, such as financial losses or physical injuries, may take many forms. AI-driven harms may handle many forms and be caused by many AI systems and applications. Some of the instances of harm caused by AI systems include:

- Self-driving car accidents: In 2018, a pedestrian in the US state of Arizona was killed by an autonomous vehicle operated by the ride-sharing technology service Uber [3]. In 2020, a Tesla owner was killed in a collision with a semi-trailer truck using Tesla's Autopilot feature [4]. Similar incidents occurred in 2022 [5].
- Healthcare AI errors: While AI is intended to reduce diagnostic errors, there is the risk that the use of AI can introduce new potential errors [6]. A 2019 analysis detailed new possible mistakes [7]. IBM's Watson also provided incorrect cancer treatment recommendations [8].
- AI-powered weapons: Concerns have been raised about the potential use of AI in lethal autonomous weapons systems, which could raise liability issues if they cause harm to civilians or military personnel, violate established rules of engagement, or commit acts in contravention of customary international law [9].
- AI-powered financial algorithms: In 2020, AI-powered trading algorithms were estimated to contribute to a dramatic drop in the value of certain stocks, leading to significant losses for investors [10], [11].
- AI-powered customer service chatbots: In 2017, an AI-powered customer service chatbot developed by a major airline made inappropriate comments to users [17].
- AI-powered virtual assistants: In 2017, a person's pet parrot fooled an Amazon Alexa virtual assistant device and made unauthorized purchases [12].
- AI-powered hiring algorithms: In 2018, an AI-powered hiring algorithm developed by Amazon showed bias against female candidates. This raised concerns about the potential for AI products to perpetuate existing biases and discrimination [13].
- AI-powered loan algorithms: In 2020, an AI-powered algorithm used by Apple showed bias against specific gender, leading to a lawsuit [14].
- AI-powered facial recognition software: In 2020, a primary tech provider's AI-powered facial recognition software had misidentified individuals [15], discriminated against Asians [16], and caused other harms [17].

- AI-powered language translation software: In 2017, a Facebook mistranslation led to the arrest of a Palestinian man [19]. In Turkish and Hungarian, AI translations are known to exemplify gender bias [20].

All these instances exhibit the potential harm caused by AI in multiple domains. To minimize harm and ensure victims may receive the appropriate compensation for their losses and damages, it is essential to understand the risks that AI systems can pose. A proper liability framework can facilitate victims' recovery of losses or damage caused by an activity, product, or event. Legislators and regulators can demand to allocate responsibility for various contributors that lead to such losses or damages based on the nature of their role in the eventual harm or damages caused by such unintended consequences of AI systems [21].

## 3 UNIQUENESS AND COMPLEXITY OF LIABILITY FOR AI-HARMS

### 3.1 Complexity of liability for AI-driven harms

While assessing liability for harm contributed to or caused by AI, it is necessary to understand how the liability in the context of AI may differ from typical product liability. AI integrates learning from data, pre-trained models, and optimization techniques adopted. The liability consideration for AI products is complex, as compared to standard product liability, due to the complexity of AI technology, unforeseen results, and the involvement of multiple economic actors and their interrelationships. Determining liability for harm, injuries, or defects caused by artificial intelligence (AI) can be difficult and complex. There are several reasons for this:

- Number of economic actors involved: AI products and services often involve multiple economic actors, such as manufacturers, developers, distributors, and end users. This can make it challenging to identify which party is responsible for a defect or harm caused by the AI.
- Level of automation of task responsibility: The level of automation relating to task responsibility is essential. In AI, determining the task responsibility associated with the potential harms is complex, and the expert or otherwise role of human-in-the-loop also needs consideration [52], [39].
- The complexity of relationships: The relationships between these economic actors can be complex and may change depending on the specific use case. For example, an AI product may be developed by one provider, manufactured by another, and sold by a third party. Each of these parties may have different roles and responsibilities, making it challenging to determine liability. There are many parties involved in the process. This makes it challenging to identify the person who caused harm. There are parties involved in creating an AI system that is stand-alone or embedded into a product, as well as parties with control over its operation. Regulators must determine how each actor behaved or their actions or inactions contributed to causing harm [1].
- Nature of tool licensing: AI products and services may be licensed to end users through various arrangements, including open-source licenses, proprietary licenses, and subscription-based models. The terms of these licenses can affect liability for defects or harm caused by AI.
- Jurisdictional differences: Different countries and regions have different laws and regulations governing product liability and AI. This can make it challenging to determine which jurisdiction's laws apply in a particular case and how those laws should be applied.
- Contractual terms: Any contracts or agreements between the economic actors involved in an AI product or service can also affect liability for defects or harms. For example, a manufacturer may include indemnification provisions in a contract with a developer to shift liability for defects or harm caused by the AI to the developer.
- Difficulty in identifying responsibility: Finally, it can be challenging to identify who is responsible for defects or harms caused by AI due to the complexity and unpredictability of the technology, as enumerated in previous sections. AI systems may be composed of many different components and rely on large amounts of data and algorithms, making it difficult to pinpoint the root cause of a defect or harm.

## 4 OVERVIEW OF THE EU LIABILITY DIRECTIVE

### 4.1 General Principles relating to Liability in the EU context

AI brings fundamental environmental changes, such as heightened complexity, worsened opacity, and increased autonomy, which can all impact liability determinations in litigation [1]. Many jurisdictions still need to address the new issues AI systems raise concerning liability frameworks. Most technological ecosystems have complexity associated with liability rules, including autonomous vehicles or programs. Hence, general tort law approaches, traditional products liability, and contractual liability may be applied, in whole, or in part, in these circumstances [1]. There is a need for legal certainty for businesses and ensuring consumers are well protected in case of AI-driven harm [22]. In September 2022, the European Commission presented two proposals. These were a new AI Liability Directive and an update of the Product Liability Directive. These regulations' substantive rules are still to be finalized. EU regulations expect the disclosure of evidence with a focus on fault, defectiveness, and causality in assessing liability [22].

Tort law focuses on civil claims that are not contractually related and arise out of incidents in which a per or suffers injury or harm caused by the act or omission of another person [23]. Generally, injury refers to the deprivation of a legal right or privilege, and harm refers to the specific loss or detriment suffered [23]. In examining a tort claim, the court will look to each party's different actions or omissions to determine how best to assign fault or responsibility for the injuries or harm caused. For instance, a tort claim alleging negligence requires the aggrieved party (often the one who initiates the civil claim) or plaintiff to demonstrate that the tortfeasor had a duty to the plaintiff, that the tortfeasor breached that duty, that the tortfeasor's breach of that duty was the cause of the harm or injury that the plaintiff suffered, and that the harm or injury caused financial losses or damages [26]. However, plaintiffs may pursue other legal theories related to their claim beyond a traditional negligence tort, such as a product liability claim.

In a typical product liability claim, the plaintiff can allege a defect in a consumer-facing item. To prevail in a products liability claim, the plaintiff must first demonstrate that there existed a defect in the manufacturing process, a flaw in the item's design [24], a misrepresentation or inadequate warning accompanying the product, [25] or a breach of a written warranty (or implied warranty) accompanying the product. [27]. A warranty, essentially, is a guarantee from the manufacturer (or seller) that the product is free from defects, will behave/function in a certain way, or be suitable for specific uses [28]. In many jurisdictions, a professional manufacturer or seller might be presumed to offer a basic product warranty unless that manufacturer or seller specifically disclaims it [28]. Depending on the circumstances, courts may also consider whether the harm or injury caused by the item was reasonably foreseeable to the manufacturer in instances where a consumer uses a pioneer beyond an item's typical use profile and whether a manufacturer's liability should be adjusted based on such an expectation.

In addition, courts may examine contract law as a basis for allowing claims between parties. Contract law, in many jurisdictions, allows for tremendous deference to the contracting parties when they have equal bargaining power concerning a negotiation. However, nearly all consumer-facing contracts offered by companies are not negotiable. Consumers call these contracts "boiler-plate", and lawyers call them "contracts of adhesion" [29]. "[C]ontracts of adhesion [are] form contracts offered on a take-or-leave basis by a party with stronger bargaining power to a party with weaker power" [30]. For decades, online consumers have been signing and clicking away their rights such as the right to a jury trial, [31], the right to bring a claim near where a plaintiff lives, [31 ], the right to sue for negligence [31 , discussing exculpatory c lauses], the right to privacy, and others [31 ]. Law Professor Margaret Jane Radin observed in 2013, "[i]n short, if you are like most US consumers, you enter into 'contracts' daily without knowing it, or at least without being able to do anything about it." [31 ].

Varying tort frameworks can allow for various claims against AI producers and companies that may use AI in their products. Courts will need to wrestle with the extent to which an AI user may have a duty toward their clients or customers [53]. Or, perhaps, in the instance of an autonomous vehicle ("AV") colliding with a non-autonomous vehicle, the courts will need to determine the factual cause of the crash (i.e., system error, operator error, negligence by the non-AV driver, etc.), and apportion responsibility between the parties involved. However, some jurisdictions have passed laws assigning liability to certain parties in certain circumstances, such as environmental protection legislation, professional licensing rules, consumer privacy, workplace safety, and employment laws. Whereas traditional tort law, product liability, and contract law, at least in common-law jurisdictions, evolve through the courts' approach, statutory liability creates specific obligations and dictates how the courts should approach them.

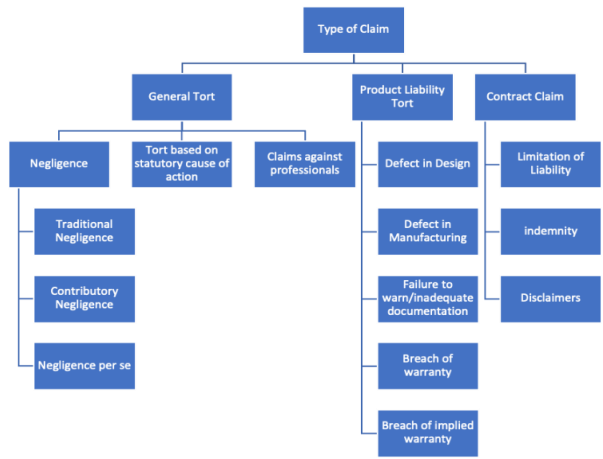

Figure 1: Illustrative overview of the different liability claims

Another matter that courts need to determine relates to the assignment of fault. While the tortfeasor may bear the primary responsibility for the chain of events leading to injury or harm, the victim may also bear some responsibility for the claim. When facing a negligence claim, some jurisdictions take an "all-or-nothing" approach wherein the tortfeasor bears all or no responsibility. However, others evaluate the extent to which the injured party bears responsibility and apply a "contributory negligence" framework wherein awarded damages may be reduced based on the extent to which the aggrieved party's actions contributed to the incident.

In determining fault, a court will look at many factors within the chain of events leading up to the event causing the injury or harm. In a well-known American case called Palsgraf, the Court of Appeals of New York heard a claim where a woman was injured by some scales that fell after a railroad employee helped a man jump onto a train as it left the station. That man dropped some fireworks that exploded, creating a shock that led to the falling of the scales [32]. A future US Supreme Court Justice, Benjamin Cardozo, authored the opinion in which the Court actually found that the railroad employee did not have a duty toward Ms. Palsgraf since the chain of events leading to her injury was not reasonably foreseeable to that employee [32]. And, while the Court ruled against Palsgraf on that basis, the reasoning by Cardozo in examining the chain of events influenced courts to consider how far back to trace a chain of events in determining fault. Finally, some tort claims must be brought within a specific time limit from when they occurred. If the jurisdiction has passed a law requiring a claim be brought within a certain period from the plaintiff's knowledge or recognition of the injury or harm, often referred to as a statute of limitations, the jurisdiction will not permit claims brought after that period. However, suppose there is no statute of limitations related to a particular claim. In that case, the courts may look to certain common-law doctrines, such as the laches, or the "[u]nreasonable delay in pursuing a right or claim", which a defendant can raise [33].

To minimize the risk to consumers (and also minimize liability), sellers and manufacturers should ensure that their products have been thoroughly tested—and include adequate instructions and warning labels. Manufacturers and sellers must keep abreast of the latest AI developments to identify and resolve potential problems before they become a problem. The future ability of the provider to contest a product liability claim will rely on good practices adopted in designing, developing, deploying, and maintaining AI systems in the market

**4.2 Salient provisions from EU liability directive**

Damages can be caused by fault, defect, failure, and adverse incident, and such injuries can lead to liability by collating evidence of such damage, associating causality to the damage, and demonstrating information inadequacy in contracts, terms of use, disclosure, warning, notification, alerts or guidance about such potential impacts.

Some salient features from the Product Liability Directive and AI Liability directive are provided below [22]:

1. The product liability Directive changes the nature of suits alleging injury or harm caused by the AI system. It also alters the evidentiary requirements for fault-based liability systems. The Directive does not create new legal obligations but could impact many claims covered by EU national law, such as harm to life, health, or property.
2. It expands the definition of Product to include intangible goods (i.e., software and AI systems). It broadens the scope of damages to include data loss and corruption (if it is not exclusively used for professional purposes). Defects now also impact the product even after deployment (for AI that continues to learn).
3. A claimant may seek an order to compel a defendant's disclosure of relevant evidence through the legal discovery process. Still, it appears that a defendant's failure to comply may lead to an abject inference: a presumption that the defect exists.
4. Regarding cybersecurity, the proposed Directive indicates that a p cybersecurity vulnerability will be considered a defect.
5. If the claimant needs help proving the defect or causation because of technical or scientific complexity, the defect/causation can be assumed based on sufficient relevant evidence.
6. An economic operator referred to in Article 7(1) may not be liable for damage caused by a defective product if that operator proves that the product was not placed on the market, put into service, or made available to the public at the time of its manufacture or sale. An economic operator may also be exempt from liability where the product's defectiveness is due to other factors, such as software updates, upgrades, or other modifications.
7. There are two defenses critical for providers: development risk defense and later defect defense. The 'development risk defence' (DRD) frees the producer from liability if, based on the state of scientific and technical knowledge at the time that the product was put on the market, it was not possible to foresee the defect. The 'later defect defense' frees the producer from liability when it is probable that the defect did not exist when the product was placed on the market but came into being afterward.
8. There will be a general statute of limitations of 3 years, so the injured or harmed party has 3 years to file a claim, while the producer cannot be held liable for damages 10 years beyond the date the product was put on the market. Further, the 10-year threshold will reset to the time of any significant modifications, including software updates.
9. Further, the Directive covers two key expectations: the presumption of causality and transparency and disclosure obligations. By creating a rebuttable presumption of causality, the claimant will need not prove the fault that caused the harm or injury, so far as the AI output resulted in harm or injury. As a result, the fault will likely rest with human conduct that developed the AI, and such behavior violates EU AI Act or harmonized standards. The providers of High-Risk AI systems are bound to disclose documentation, logging, record-keeping, risk assessment, and transparency obligations under EU AI Act to the claimant if the high-risk system results in a fault-based liability claim.

While there are concerns and commentary around later defense, material defect definitions, and other legal and practical considerations [34], [35], the need to establish a sufficient mechanism to handle liability for AI harms is becoming certain.

## 5 OVERVIEW OF KEY APPROACHES TO LIABILITY AND THEIR LIMITATIONS

### 5.1 Key approaches to liability

Regarding AI incidents, scholars advocate for different liability regimes [36]. Liability research in the case of AI harms can generally be based on (1) tort law and (2) contract law. In some EU jurisdictions, contract liability is extended under certain conditions allowing a third party to invoke a contract to which they were not a party

themselves – in essence, allowing such parties to become what a lawyer might call a "third-party beneficiary" [1].

In many cases, AI liability is viewed from the lens of product liability. As discussed above, product liability can take many forms and remains essential in artificial Intelligence (AI). AI products can have unanticipated consequences that cause harm or injury both to users and third parties. Research also suggests the need for supplemental rules (industry, domain, or use case specific) to complement better the legal premise regarding liability determination [37].

One such suggestion is applying the enterprise theory of liability. Enterprise liability makes the members of a shared enterprise jointly and severally liable for harms caused by AI systems meaning that a plaintiff can recover damages from either or both parties. Joint enterprise liability helps determine, facilitate, and apportion responsibility among its members [38]. These approaches understand that exposing developers to strict liability will result in greater social welfare. Developers' strict liability should be limited to the risk that can be perceived based on state of the art [1]. The operator's liability will be determined based on the severity of risk, heterogeneity of operators, and the level and degree of automation. Front-end operators will have fault-based liability, and there will be strict liability for back-end operators [1]. Some research also points to a vicarious liability approach for determining liability for AI harms; however, such an approach is complex due to the need to establish a relationship of control or supervision between the party being held liable and the party whose actions or omissions caused the harm [41]. This concept again expands onto the product liability perspective using tort law.

### 5.2 Limitations of some of the existing approaches

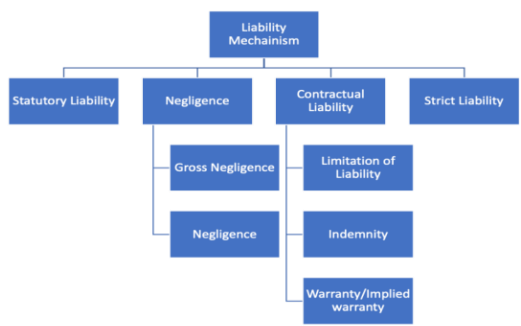

Figure 3 : Illustrative representation of various liability streams

#### 5.2.1 *Negligence*

A negligence approach would require consideration of whether the AI, its operator, or its developer had a legal obligation or duty, whether they breached that duty, whether there was an injury or harm, and whether causation could be established between the breach of duty and the injury or harm. Establishing this is tough in many cases owing to the complexity of determining user conduct. AI errors may be unforeseeable owing to the opacity of the computational models used to generate decisions or recommendations, making it challenging to determine a breach of duty. Accountability changes when the AI system is opaque, leading to the need for a more precise method of allocating responsibility [46]. While the standard of care regarding the proper use of AI systems exists, attempts associated with moral engineering for appropriately balancing AI's potential benefits and risks of harm may not be sufficient [45]. For instance, the efforts towards automated censorship and automated content moderation (engineering morality) have received high levels of scrutiny.

### 5.2.2 *Duty of care*

For negligence claims, courts must determine whether a defendant had a duty of care concerning the plaintiff. Courts use the defendant's duty of care to determine liability. The duty of care can rely on proximity, foreseeability, fairness, and concepts of reasonableness in determining liability for artificial intelligence (AI) harms [39]. Foreseeability relates to the ability to predict the possibility and likelihood of a specific harm or injury when the AI system was created or sold. To establish such liability for AI harm or damages, the victim must prove that the harm or injury was predictable and that the provider should have prevented it. Proximity refers to the relationship between parties in an AI system and the harm caused. To establish liability for AI harm, the plaintiff must establish sufficient duty and proximity between the responsible person and the victim. Further, the notions of fairness, justness, and reasonableness apply not only to the defendant's actions but also to a court's prescribed remedy. This considers the relative contributions of each party to the harm or injury and their ability to prevent or mitigate it.

### 5.2.3 *Strict Liability*

Strict liability, as often applied to abnormally dangerous activities, such as transporting hazardous waste, is another avenue for liability [47]. The challenge is that strict liability requires the owner to have control or the ability to control the AI causing harm [48]. However, this approach becomes limited since the owner may sometimes need more control over AI systems.

### 5.2.4 *Defect and failure*

Under the product liability approach, the aggrieved injured party can claim damages caused by defective design, error in manufacturing or failure to warn [42] [43] adequately. There are two primary obstacles in applying the product liability approach to AI: the effort to prove defect (even causal impact) and the enterprise liability doctrine. It states that the manufacturer is not obligated to share the warning with patients if they have shared it with the physicians [38]. It is pertinent to note that AI-related risks are not always foreseeable by the manufacturer since they may emerge after the AI is deployed in the market. Further, Product liability may not cover indirect harm with considerable disadvantages to the person concerned [44]. For instance, even the best AI in the medical industry may cause iatrogenic injury and damage, resulting in potential harm [45]. It is also limited as it covers only the damages caused to the health and property.

Similarly, in corporate law, the so-called agency issue holds the principal liable for their agent's actions. Agency is AI systems capability and precautions defined by upstream designers and deployers to limit the downstream users from expanding or changing such capabilities. It depends on the principal's supposed controls, oversight, and supervision of the agent and ability to modulate the actions. Given the multiple actors involved in building an AI system, it is challenging to determine which actors had control, oversight, or supervision over the said scenario that resulted in harm.

### 5.2.5 *Guarantee and Warranty*

Some research suggests the concept of product guarantee as an alternative to address the challenges associated with liability [36]. However, even for establishing such a mechanism, liability risk must be assessed. Similarly, some research suggested adopting an approach that determines liability for harms contributed by each stage of the AI design, development, and deployment process [52]. However, such an approach requires deterministic outcomes on the root causes of the liability. Determining root causes (associated with stages of development) for AI harm is an open issue in many cases of AI harm.

For instance, Watson for Oncology recommended a drug called Taxane for cancer that has not yet spread to the lymph nodes, which is usually prescribed only when cancer has spread to the lymph nodes [51]. In specific instances where the AI system is deployed in high-risk contexts (surveillance, recidivism, cancer treatment, etc.), there are difficulties in determining the actor who is the primary cause of the harm, given the involvement of multiple actors [51].

### 5.2.6 *Performance or safety*

Establishing responsibility in the context wherein no universally or industry-wide accepted standards for performance or safety/security for such an AI system exists is challenging. In such cases, some jurisdictions apply the Bolam Test . It states that a doctor cannot be negligent if their actions would be considered reasonable by an appropriate body of medical opinion in that specialty at the time [53]. Determining reasonable efforts, controls, and risk mitigations is one of the critical steps in determining liability. In such cases of a liability claim, it is necessary to determine what is reasonable and what a reasonable user may do—determining the essential information will help them to know reasonable information regarding the AI for the anticipated action or potential liability.

### 5.2.7 *Determining Causation*

While there are attempts to bring the presumption of causality to apportion the liability and to address the difficulties experienced by victims in explaining in detail how fault or omission caused harm in the context of an AI system, it still requires a methodical approach to specific issues contributing to a liability claim [22]. Causation arrives when the harm would not have occurred but for the conduct/risk ('but for test') relating to the AI product owner [48]. This also requires detailing conditions under which the requirement for the burden of proof will be reversed for the existence of a fault in the AI system [44]. Further, adversary systems may manipulate inputs to misclassify; however, the owner of the AI product may not have ways to identify or detect the input defects [48].

### 5.2.8 *Contractual Liability*

Many jurisdictions offer great deference to contracting parties to establish the terms for the contract, including elements such as remedies, liability, applicable law, jurisdiction, venue, and terms of performance.  As a result, courts will often permit contract terms to stand if challenged—unless such terms are illegal, unlawful, impossible to perform, or if the terms are drawn so favorably to one party and/or such a disparity between the parties exists that a court may invalidate specific terms in a contract as unconscionable.  However, courts have often permitted one-sided consumer-facing contracts [31].

### 5.2.9 *Statutory Liability*

In some cases, such as using an AI system by a state, provincial, or local government, the harm caused by AI systems may be widespread and affect many people or organizations. In these situations, specific methods may be necessary for allocating risk. This could involve implementing measures to mitigate the risks associated with the AI system or distributing the costs of any harm among a larger group of stakeholders.

### 5.2.10 *Liability in cases where experts are in loop*

AI where experts are in the loop – experts may be responsible for harm caused if they could anticipate the fault or failure and override the system [39]. This is specifically in cases where a physician relies on AI for clinical decision-making. Research suggests that relying on AI for non-standard penalties may increase the liability risk of physicians [49]. While there are threats that AI brings in automation bias in the minds of experts in the loop, human-in-the-loop cannot be made liable for harm unless they have sufficient opportunity, subjective experience, and capability to override the AI system.

For instance, if a radiology interpretation provides only its interpretation of the images but does not provide source radiology image access to the doctors, the experts in the loop do not have the opportunity to identify inconsistencies.

### 5.2.11 *Explainability*

Interpretation defects can also emerge due to a need for more explainability. Explainability has a two-fold problem: (1) victim's or users' incorrect interpretation of the explanations provided by the system; and (2) inconsistent explanations provided by the AI system. The liability in the former may be attributable to the user's sphere of events subject to certain conditions; the liability for later clearly falls on the provider (or developer of the AI system). AI liability cannot be determined by the explanations provided by the AI system, given the unreliability of these explanations [50].

## 6 RISK-BASED APPROACH TO LIABILITY

### 6.1 Need for a risk-based approach to liability

Deterrence, risk acceptance, and economic incentives are key factors in determining optimal liability. While regulators aim to develop more robust liability directives, examining liability risk is becoming paramount. The risk-based approach enables the assessment of preventable and residual risks and helps develop mitigation strategies for preventable risks. It supports transparency and disclosure mechanisms for minimizing residual risk impact. This paper provides key considerations for assessing the risk of liability for AI with the EU's AI liability directive and Product Liability Directive in the context of this section.

A liability risk assessment identifies and analyzes the potential risks and liabilities associated with a product or activity. This can help businesses understand potential risks and liabilities and develop strategies to manage those risks. Businesses can use liability risk assessment to help them identify possible risks and liabilities they might not have known about. Assessing liability risk can help companies determine the likelihood and severity of the risk. This can help prioritize risk management efforts such as expanding insurance coverage or taking risk-mitigation steps.

### 6.2 Approach toward liability risk assessment

The accuracy of risk analysis depends on the quality of the risk inputs. Risk analysis is formulated based on the likelihood of harm and the seriousness or impact of such harm to people, community, and nation/ state, caused by the autonomous decision. The provider can determine the overall risk level by conducting a risk analysis and defining the appetite and tolerance level for liability in metrics, thresholds, and benchmarks by documenting the facts relating to each question and assigning a risk rating for each question.

The questions are divided into multiple sections, namely, (A) Regulatory compliance, (B) Liability assessment mechanism, (C) Pre-deployment testing mechanism, (D) Post-hoc testing mechanism, (E) Transparency and Disclosure, (F) Adverse incident reporting system, (G) Crisis management and harm prevention, (H) Dealing with third parties in the supply chain, and (I) Risk management mechanism. These questions are created as a self-assessment tool to determine the liability risk level of an AI system. The risk assessment shall be consolidated at a section level to see whether they have many questions representing a specific risk level (High/ Medium/ Low).

Risks shall be classified as high, medium, and low based on the following principles:
1. To what extent will the results affect the user's life?
2. To what extent will the outcomes affect users' rights and freedom?
3. To what extent is it intend to uphold ethical principles?

A guide for determining risk consideration for each of the questions is provided below [54]:

| For High Risk | For Medium Risk | For Low Risk |
|---|---|---|
|  |  |  |
| - AI results may have a bearing on the safety and security of individuals or may determine their critical life decisions (e.g. disease diagnosis algorithms).<br>- AI is anticipated to impact individuals' eligibility for certain benefits, thereby significantly impacting rights or freedom.<br>- AI aims to expand existing societal biases. | - AI results may affect individuals' convenience or financial choices.<br>- AI is anticipated to impact individuals' eligibility for certain benefits to a limited extent due to public interest requirements.<br>- AI intends to support determining prioritization for essential services. | - AI results may have limited bearing on individuals and may not impact them physically or financially.<br>- AI intends to limit potential access to services or opportunities (social or technical accessibility). |

### 6.3 Liability risk self-assessment questionnaire

#### 6.3.1 *Regulatory compliance*

1. Nature of activity: Whether the activity AI performs can be construed to be illegal or unlawful? Is the AI exposed to potential attacks that may make it perform illegal or unlawful activities?
2. Safety regulations: Is the provider required to follow any safety regulations for its AI product or service? Are there indications of not complying or deficiencies in compliance that can lead to a liability claim?
3. Regulatory compliance: Does the provider understand the laws and regulations governing product liability and AI in the jurisdictions where the AI product will be sold?
4. Certification mechanism: Does the provider have a mechanism to be certified or approved by any regulatory bodies before selling it? Does the regulation mandate this?
5. Harm definition: Does the product have an established definition of harm? Does the definition of harm include tangible and digital harms?
6. Differential sources of harm: Does the provider have adequate mechanisms to differentiate the source of harm into design defects, development defects, or defective warning or safety instructions?
7. Third-party impact assessment: Does the provider have a mechanism to assess if the AI product could pose any potential dangers to third parties and take steps to mitigate those risks?

#### 6.3.2 *Liability environment*

1. Policy review: Does the provider regularly review, and update policies and procedures related to AI product liability to ensure that they are adequate and up to date?
2. Liability or compensation schemes applicability: Are some of AI's harms also perceived to fall under the non-fault compensation schemes (for reasons including inability to identify tortfeasor) as may be applicable under relevant regional regulations?
3. Mandatory insurance: Does the provider industry or the use case fall under any form of mandatory insurance scheme for potential liability? How is the provider determining the extent of insurance coverage required?
4. Indirect damages: Has the provider assessed the potential for liability due to indirect damages contributed by AI harms?
5. Industry threat monitoring: Do providers monitor industry developments and research related to AI product liability to stay current on potential risks that can lead to liabilities?
6. Liability handling mechanism: Whether the provider has an established process for handling and responding to product liability claims that may arise?
7. Product recall mechanism: Does the provider have a structured plan for recalling the AI product in case of a defect or malfunction?
8. Industry benchmarking: Has the provider conducted performance benchmarking against industry benchmarks? Whether the performance meets the duties of care of the operator?

### 6.3.3 *Pre-deployment testing mechanism*

1. Reliability testing: Did the provider test the AI product to ensure it is fair, reliable, and safe?
2. Causality testing: Does the provider test for potential causality failures (factual and legal) of the self-learning or adaptive AI system that may lead to liability?
3. Historical data: Does the provider have a list of liability claims and complaints based on harms contributed by similar systems compiled through secondary research and market assessment?
4. Open source: Does the provider work with any open-source tool or model or service provider in exchange for a fee or the exchange of personal data?
5. Cyber threats: Does the provider have measures to protect and detect potential cyber threats to the AI product?

### 6.3.4 *Post-hoc analysis mechanism*

1. Post-hoc analysis: Does the provider deploy measures to conduct post-hoc analysis (traceability, accessibility, intelligibility, and comprehensibility) of the events relating to adverse incidents reported or learned by it?
2. Explainability: Does the provider have a sufficient explainability mechanism to demonstrate that different input factors resulted in different or similar results and vice versa?
3. Cybersecurity attack: Does the provider have a mechanism to assess the harms contributed by the exploitation of AI by hackers or attackers or similar other groups that have the intent to cause damage to the AI or its users?
4. Conduct of user: Does the provider have a mechanism to assess the harms contributed by the victim/claimant's sphere or associated conduct or behavior?
5. User behavior in the AI system: Does the provider have a mechanism to track the user's behavior (via logs) within the product or application to determine if the user's conduct led to harm?
6. Quality testing: Does the provider test the quality of the product to assess the potential damages it may cause when deployed in the market?
7. Root cause analysis: Does the provider have sufficient logging and disclosure mechanism to trace root causes of potential harms?
8. Testing mechanism: Does the provider have tests, measures, and metrics to track product overuse/ misuse/ abuse?
9. Performance monitoring: Does the provider implement a process for tracking and recording any defects or harms based on performance monitoring of the AI product?
10. Safety monitoring: Does the provider conduct an adequate monitoring mechanism around the AI product's design, development, and deployment to ensure that the product is not inherently unsafe or causes harm to the users?
11. Post-market monitoring: Does the provider have any mechanism to adequately monitor the AI after putting it into circulation (or the market deployment)? Does the provider have a mechanism to track potential performance deterioration on an ongoing basis that may cause harm to users?

### 6.3.5 *Transparency and disclosure*

1. Contractual relationship with the user: Does the provider have an established contract mechanism clarifying the user's and product's roles?
2. Potential use information: Does the provider provide adequate disclosures regarding the product's potential uses that discourages the users? Does the provider have a process for educating users about the proper use and care of the AI product?
3. Instruction of use: Does the provider provide instructions for the use of the product and limitations regarding the use of the product?
4. Marketing ethical use: Does the provider design, describe, and market products in a way that effectively enabling users to understand the appropriate/eligible uses of the AI?
5. Capability information and limitations: Does the provider provide clear and concise documentation explaining the AI product's capabilities and limitations to help users understand its potential risks and benefits?
6. Warning labels: Does the provider provide clear and accurate warning labels and instructions to help users understand the potential risks associated with the AI product?
7. User notification mechanism: Does the provider have a mechanism of notifying the users, warning them of potential risks associated with the use of the product as they become aware?
8. Log collection and maintenance: Is the AI product designed with a "logging" & "debugging" feature to track and analyze any defects or malfunctions?
9. Provide access to the victim: Does the provider provide adequate access to the logs or otherwise provide access to information for the victim to establish a liability claim?

### 6.3.6 *Dealing with third parties*

1. Stakeholder map: Does the provider have an established mechanism for documenting the hierarchy of relationships between stakeholders involved in the final product?
2. Contractual arrangement: Does the provider have adequate contractual terms for binding the other economic actors / contracted third party into a liability for harm?
3. Warranties: Does the provider require the other economic actors involved in the AI product to provide warranties for defects?
4. Liability clause: Does the provider have appropriate limitation of liability clauses for the failure or defect of other economic actors involved in the AI product?
5. Indemnification clause: Does the provider have adequate indemnification provisions or limitations on liability included in contracts with developers, manufacturers, or other economic actors involved in the AI product?
6. Liability attribution: Does the provider have sufficient measures to attribute liability to specific ic economic actors based on assessed levels of harm?
7. Fault identification: Does the provider have a mechanism to identify the risks of fault or defect relating to the final product? Does the provider determine the likelihood of such faults?
8. Root cause analysis: Does the provider have a mechanism to determine the potential liability arising due to technical interdependency and interoperation with tools or technology relating to other economic actors or the degree of specificity or exclusivity of their relationships?
9. Defect reporting: Does the provider have an established process for handling and reporting defects or malfunctions that occur with the AI product?

10. Complaint management: Does the provider have an established process for receiving and responding to customer feedback and complaints about the AI product? Investigating adverse incidents: Does the provider have a mechanism for root cause analysis and investigations of reported adverse incidents?

### 6.3.7 *Crisis management and harm prevention*

1. Back up mechanism: Is the AI product designed with a failsafe or backup system to prevent/minimize harm in the event of a malfunction?
2. Kill switch: Does the AI product have a "kill switch" or another mechanism to be used to shut down the AI system in the event of a malfunction?
3. Remote Shutdown: Does the AI product have additional safety measures or controls, such as remote shutdown capabilities or automatic updates?
4. Human interventions: Is the AI product designed with a "human in the loop" or another mechanism low for human intervention in the event of a malfunction?

### 6.3.8 *Risk Management*

1. Risk assessment for potential misuse of AI system: Does the provider have a risk assessment mechanism to identify possible uses of the product or scenarios wherein the use of the product may be unsafe or cause harm to the users?
2. Risk review for defects and failures: Does the provider conduct a risk review of potential product defects and failures that may result in liability as applicable to the relevant laws and regulations governing product liability and harms caused by products?
3. Insurance review: Do the provider review and update product liability insurance to ensure that it adequately covers the risks associated with the AI product?
4. Residual risk management: Does the provider have sufficient measures for dealing with residual risks that may result in harm? Legal defenses for liability: Does the provider have a sufficient legal defense for potential liability due to defects or harm caused by the AI product?

## 7 CONCLUSION

Since the providers are unaware of all the risks associated with the unintended consequences of AI systems, they remain unaware of their potential liability. This questionnaire provides them with an initial assessment process of providing a detailed approach to assessing liability risk for their products. It also provides an opportunity to shed light on risks relating to AI systems to allow for more pointed and serious conversations around the issue. After all, as Justice Louis Brandeis observed, "[ s]unlight is said to be the best of disinfectants" [40].

Providers can take many steps to reduce or mitigate risk by understanding where their actions create or enter risk in the development. However, it is necessary to conduct an adequately detailed risk assessment to understand the underlying factors that could result in harm. Using such a questionnaire or preliminary assessment supports identifying risks and developing monitoring programs to mitigate potential downstream injuries and harms, potential financial liability and exposures, and potential reputational harm. Such mechanisms also help governments better understand enterprises and produce more effective regulation. The questionnaire provides preliminary direction to the liability risk assessment; such a self-assessment cannot be adopted in isolation without adequate risk and compliance mechanisms concerning AI systems. Also, the self-assessment questionnaire does not support quantifying liability; it only supports assessing the level of liability risks. This preliminary questionnaire is intended to provide direction for further research and development in assessing and quantifying the potential liability.

[51] IBM pitched Watson as a revolution in cancer care. It's nowhere close. (n.d.). Retrieved December 31, 2022, from https://www.statnews.com/2017/09/05/watson-ibm-cancer/

[52] Lagioia, F., & Contissa, G. (2020). The strange case of dr. Watson: Liability implications of ai evidence-based decision support systems in health care. European Journal of Legal Studies, 12(2), 245–289. https://doi.org/10.2924/EJLS.2019.028

[53] Test of Medical Negligence. (n.d.). Retrieved December 31, 2022, from http://www.legalservicesindia.com/article/1685/Test-of-Medical-Negligence.html

[54] Fairness Assessment and Rating of Artificial Intelligence Systems. (n.d.). Retrieved January 8, 2023, from www.tec.gov.in
17